\documentclass[prl,twocolumn,nofootinbib,aps]{revtex4}
\pdfoutput=1
\usepackage[pdftex]{graphics}
\usepackage{graphicx}
\usepackage{epstopdf}
\usepackage{amsmath}
\usepackage{amsfonts}
\usepackage{bm}
\usepackage{epsfig}
\usepackage{graphics}
\usepackage{xspace}
\usepackage[usenames]{color}
\newcommand{\roughly}[1]{\mathrel{\raise.3ex\hbox{$#1$\kern-0.85em
\lower1ex\hbox{$\sim$}}}}

\def\bea{\begin{eqnarray}}
\def\eea{\end{eqnarray}}
\def\be{\begin{equation}}
\def\ee{\end{equation}}

\def\nn{\nonumber}

\def\({\left(}
\def\){\right)}

\begin{document}

\newcount\hour \newcount\minute
\hour=\time \divide \hour by 60
\minute=\time
\count99=\hour \multiply \count99 by -60 \advance \minute by \count99
\newcommand{\mydate}{\ \today \ - \number\hour :00}

\preprint{???}
\preprint{Pi-Partphys-???}

\title{On theories of enhanced CP violation in $B_{s,d}$ meson mixing.}
\author{
Michael Trott${}^1$ and Mark B. Wise${}^2$\\
${}^1$ Perimeter Institute for Theoretical Physics, Waterloo ON,
N2L 2Y5, Canada.\\
${}^2$ California Institute of Technology, Pasadena, CA, 91125, USA, }

\date{\today}

\begin{abstract}
The $\rm DO \! \! \! \!/ \, $ collaboration has measured a deviation from the standard model (SM) prediction in the like sign dimuon asymmetry in 
semileptonic $b$ decay with a significance of $3.2 \, \sigma$.  We discuss how minimal flavour violating (MFV) models with multiple
scalar representations can lead to this deviation through tree level exchanges of new MFV scalars.
We review how the two scalar doublet model can accommodate this result and discuss some of its phenomenology.  Limits on electric dipole moments suggest that  in this model the coupling of the charged scalar to the right handed $u$-type quarks is suppressed while its coupling to the $d$-type right handed quarks must be enhanced. We construct an extension of the MFV two scalar doublet model where this occurs naturally.
\end{abstract}
\maketitle

\bigskip

\section{Introduction}
The $\rm DO \! \! \! \!/$ collaboration has reported a $3.2 \, \sigma$ deviation from the standard model (SM) prediction of the like sign dimuon asymmetry in 
semileptonic $b$ decay \cite{Abazov:2010hv}.  This observation joins past anomalous measurements of $B_s \rightarrow J/\psi \, \phi$ and $B^- \rightarrow \tau \, \nu$
decays that can be interpreted as a pattern of deviations consistent with new physics contributing a new phase in $B_{s,d}$ mixing (for a recent global fit and discussion see \cite{Lenz:2010gu}).\footnote{The observed $2.6 \, \sigma$ deviation from the standard model (SM) expectation \cite{Lenz:2010gu} in the averaged measurements of $B^- \rightarrow \tau \, \nu$ performed at Belle and Babar \cite{Ikado:2006un,Aubert:2007xj,:2008gx,:2008ch}
correlates correctly with a new physics (NP) contribution of a phase to $B_d$ with a sign consistent with the NP phase implied by the $\rm DO \! \! \! \!/ \, $ dimuon measurement.
Such a NP phase also correlates with the expectation of a shift in 
$\sin 2 \, \beta$ extracted from $B_s \rightarrow J/\psi \, \phi$ compared to the SM expectation  \cite{Aaltonen:2007he,:2008fj} and extractions from measurements in $B_s \rightarrow J/\psi \,  K_s$.
Such a consistent deviation is also observed, its statistical significance is $2.1 \, \sigma$. Also see \cite{Lunghi:2008aa} for a discussion on the evidence for a NP phase in $B_d$ and $B_s$ meson mixing.}

If the explanation of the like sign dimuon asymmetry measurement and these correlated deviations are not statistical fluctuations, 
then new physics interpretations of this pattern are of interest. General operator analyses have been carried out \cite{Ligeti:2010ia,Blum:2010mj,Batell:2010qw} and indicate that operators induced by scalar exchange with  
unenhanced Yukawa couplings and order one parameters in the potential (i.e. order one Wilson coefficients) the mass scale suppressing the operators of interest is a few hundred $\rm GeV$.

Such a low mass scale is challenging to reconcile with known constraints from flavour physics unless minimal flavour violation (MFV) \cite{Chivukula:1987py,Hall:1990ac,D'Ambrosio:2002ex} is imposed.
New physics (NP) models with MFV have the quark flavor group $SU(3)_{U_R} \times SU(3)_{D_R} \times SU(3)_{Q_L}$ only broken by the Yukawa couplings. However this scenario does allow new phases and so provides a framework for explaining the anomalies mentioned above without giving  rise to  flavor changing neutral current effects that are in conflict with experiment.

In this paper we discuss  scalar models with MFV that can explain these anomalies.
We first review how tree level exchanges of a neutral complex scalar in a simple two scalar doublet model can lead to enhanced CP violation in the $B_q$ meson system and discuss the phenomenology of this model. 
We then show that limits on electric dipole moments suggest  that the coupling of the charged scalar  to the right handed $u$-type quarks is suppressed while its coupling to the $d$-type right handed quarks must be enhanced to be consistent with the data. We construct an extension of the MFV two scalar doublet model  where this occurs naturally\footnote{Of course models with scalar doublets that are not supersymmetric suffer from the well know naturalness problem of keeping the doublets light compared to the Planck scale.}.

\section{Set up}

We will utilize the recent fit of \cite{Lenz:2010gu} to determine the new contribution
to $B_q-{\bar B}_q$ mixing (here $q = {s,d}$). This fit is consistent in its conclusions with an earlier analysis \cite{Ligeti:2010ia}.
The $\rm DO \! \! \! \!/ \, \, $  result ($a_{SL}^b$) and the SM prediction \cite{Lenz:2010gu} ($A_{SL}^b$) are given by
\bea
a_{SL}^b &=& \frac{N_b^{++} - N_b^{--}}{N_b^{++} + N_b^{--}}, \nn \\
 &=& - (9.57 \pm 2.51 \pm1.46) \times 10^{-3},  \\
A_{SL}^b &=& (-3.10^{+0.83}_{-0.98}) \times 10^{-4}.
\eea
where the number of $X \, b \, \bar{b} \rightarrow \mu^+ \, \mu^+ \, Y$ events is given by $N_b^{++}$ for example.
The quoted $a_{SL}^b$ is a combination of the the asymmetry in each $B_q$, denoted  $a_{SL}^{bq}$. Each of these contributions to $a_{SL}^b$ can be expressed in terms of the 
mass and width differences ($M_{12},\Gamma_{12}$) of the $B_q$ meson eigenstates and the CP phase difference between these quantities $\phi_q$  as
\bea
a_{SL}^{bq} = \frac{|\Gamma_{12}^q|}{|M^q_{12}|} \, \sin \phi_q.
\eea
Naively one can effect the SM prediction through modifying $M_{12}$ or $\Gamma_{12}$ and 
both approaches have been explored in the literature. Modifying the decay width significantly \cite{Dighe:2010nj,Bauer:2010dg}
as an explanation is problematic\footnote{The decay width can be removed in the relation between measured quantities
under the assumption of small CP violation in NP induced tree level decays of $B_q$ \cite{Grossman:2009mn,Ligeti:2010ia}
and the anomalous measurements can still be fit to finding $\sim 3 \, \sigma$ evidence for a deviation from the SM \cite{Lenz:2010gu,Ligeti:2010ia}.
Also, an explanation of the like sign dimuon asymmetry through
a NP contribution to $|\Gamma_{12}^q|$ would not necessarily explain the anomalies in 
$B_s \rightarrow J/\Psi \, \phi$ and $B \rightarrow \tau \, \nu$ with the correct correlation.} and we will focus on MFV NP
explanations that involve a NP contribution (that includes a new CP violating phase) to $M^q_{12}$. 

The effect of NP on $B_s$ and $B_d$ mass mixing can be parametrized by two real parameters,  $h_q>0$ and $\sigma_q$ by writing
\bea
M_{12}^q = \left(M_{12}^q\right)^{\rm SM}+ \left(M_{12}^q\right)^{\rm NP}, 
\eea 
where the new physics contribution to the mass mixing  is related to the standard model value of the mass mixing by,
\bea
\label{hsigma}
\left(M_{12}^q\right)^{\rm NP}=\left(M_{12}^q\right)^{\rm SM} \,  h_q \, e^{2 \, i \, \sigma_q} .
\eea
The models we discuss have  $h_s = h_d$ and $ \sigma_s = \sigma_d$ which is generally expected in NP models
that obey MFV. \footnote{It has been proven in \cite{Kagan:2009bn} that new CP violating effects can be larger in $B_s$ than in $B_d$ in nonlinear MFV. This observation has recently been explored in a general operator analysis \cite{Batell:2010qw} which showed that 
enhancements of CP violation in $B_s$ mixing over $B_d$ mixing by $m_s/m_d$ requires contributions in the 
MFV expansion out to forth order in both the up and down Yukawas for operators induced by scalar exchange. The results on the neutron EDM using naive dimensional analysis (NDA) on page 4 disfavour
order one down and up Yukawas with CP violating phases for these operators, so such contributions are expected to be very small. For alternative estimates of
the relevant matrix element not using NDA see \cite{Demir:2002gg}.}
This scenario is argued to be a better fit to the current data then the SM in \cite{Lenz:2010gu}, which is disfavoured with a p-value
of $3.1 \sigma$. In this case, the best fit values  are $h_q =0.255$ and $2\sigma_q=180^{o}+63.4^o$. The best fit magnitude of the correction $h_q$ is small but its phase  is large.

For simplicity in this paper we treat perturbative QCD in the leading logarithmic approximation and evaluate the needed matrix elements of four quark operators using the vacuum insertion approximation at the bottom mass scale. At the $t$-quark mass scale, in the SM, the effective Hamiltonian for $B_q-\bar B_q$ mixing is,
\begin{equation}
{\cal H}^{\rm SM}_q=  (V_{tq}^\star \, V_{tb})^2 C^{\rm SM}(m_t) {\bar b}^{\alpha}_L \gamma^{\mu} q^{\alpha}_L{\bar b}^{\beta}_L \gamma_{\mu} q^{\beta}_L,
\end{equation}
where $\alpha$ and $\beta$ are color indices and
\begin{equation}
C^{\rm SM}(m_t) =  \frac{G_F^2}{4 \, \pi^2} \, M_W^2 \,  S(m_t^2/M_W^2).
\end{equation}
Here  $S(m_t^2/M_W^2) \simeq 2.35$ is a function of $m_t^2/M_W^2$ that results from integrating out the top quark and $W$-bosons. Using
\begin{equation}
\left(M_{12}^q\right)^{\rm SM}={\langle  B_q|{\cal H}^{\rm SM}_q|{\bar B}_q\rangle \over 2 m_{B_q}},
\end{equation}
we have after running down to the $b$-quark mass scale that,
\begin{equation}
\left(M_{12}^q\right)^{\rm SM}= (V_{tq}^\star \, V_{tb})^2C^{\rm SM} (m_t) \left({1 \over 3}\right)\eta f_{B_q}^2 m_{B_q}.
\end{equation}
Here $\eta \simeq 0.84$ is a QCD correction factor,  $C^{\rm SM}_q(m_b) =\eta  \, C^{\rm SM}_q(m_t)$.

The models for new physics we discuss generate the effective Hamiltonian at the top scale\footnote{For QCD running we don't distinguish between the top scale, weak scale and the mass scale of the new scalars we shall add.}
\begin{equation}\label{NPH}
{\cal H}^{\rm NP}_q\simeq (V_{tq}^\star \, V_{tb})^2C^{\rm NP}(m_t) {\bar b}^{\alpha}_R q^{\alpha}_L{\bar b}^{\beta}_R q^{\beta}_L.
\end{equation}
Running down from $m_b$ operator mixing induces the  analogous operator with color indices rearranged. However, its coefficient is very small and we neglect it resulting in the relation, $C^{\rm NP}(m_b) \simeq \eta' C^{\rm NP}(m_t)$, where $\eta'  \simeq 1.45$  \cite{Bagger:1997gg}. Again using the vacuum insertion approximation at the $b$-quark mass scale we arrive at,
\begin{equation}
\left(M_{12}^q\right)^{\rm NP}\simeq(V_{tq}^\star \, V_{tb})^2C^{\rm NP}(m_t)\left(-{5 \over 24}\right) \eta'  f_{B_q}^2 m_{B_q}.
\end{equation}
Comparing with  Eq.~(\ref{hsigma})
\begin{equation}\label{key}
 h_q \, e^{2 \, i \, \sigma_q} \simeq-{5 \over 8}\left({C^{\rm NP}(m_t) \over C^{\rm SM}(m_t) }\right) {\eta' \over \eta}.
 \end{equation}

\section{Minimal Two Scalar Doublet Model}

We now discuss how the minimal two scalar doublet model with MFV can have enhanced CP violation in $B_q$ mixing due to tree level exchange of neutral scalars.\footnote{Previous analyses focused on scalar exchange to explain the like sign dimuon asymmetry include  \cite{Dobrescu:2010rh,Buras:2010mh,Buras:2010zm}. Also see \cite{Pich:2009sp,Jung:2010ik} for some phenomenological studies of models of this form.} 
We denote by $H$ the doublet that gets a vacuum expectation value and by $S$ the doublet that does not.  The Lagrangian in the Yukawa sector is
\bea\label{general}
{\cal L}_Y&=& {\bar u}^i_{R}  \, g^{~~j}_{U~i} \, Q_{L j} H+ {\bar d}^i_{R} \, g^{~~j}_{D~i} \, Q_{L j} H ^{\dagger} \,  \\
&\,& +  {\bar u}^i_{R}  \, Y^{~~j}_{U~i} \, Q_{L j} S +  {\bar d}^i_{R}  \, Y^{~~j}_{D~i} \, Q_{L j} S^{\dagger}+{\rm h.c.} \nn
\eea
where flavour indicies $i,j$ are shown and color and $ SU(2)_L$ indices have been suppressed.  
{\rm MFV} asserts that any {\rm NP} also has the quark flavour symmetry group only broken by insertions proportional to Yukawa matrices  so that 
$Y^{~~j}_{U~i}, Y^{~~j}_{D~i}$ are proportional to  $g^{~~j}_{U~i}, g^{~~j}_{D~i}$. 
One can construct allowed NP terms by treating the Yukawa matrices as spurion fields that transform under flavour rotations as,
\begin{equation}
g_U \rightarrow V_U \, g_U \, V_Q^{\dagger},~~~~~~~~~~g_D \rightarrow V_D \, g_D \, V_Q^{\dagger},
\end{equation}
where $V_U$ is an element of $ SU(3)_{U_R}$, $V_D$ is an element of $ SU(3)_{D_R}$, and $V_Q$ is an element of $ SU(3)_{Q_L}$, i.e.,
the Yukawa matrices  transform as $g_U \sim ({\bf 3},{\bf 1}, {\bf {\bar{3}}})$ and  $g_D \sim ({\bf 1},{\bf 3},{\bf {\bar{3}}})$ under the flavour group.
MFV can be formulated up to linear order in top Yukawa insertions, or extended to a nonlinear representation of the symmetry \cite{Feldmann:2008ja,Kagan:2009bn}. 
For enhanced CP violation in $B_q$ mixing we are interested in a nonlinear realization of MFV. It is sufficient to only expand to next order in insertions of $g_U$ so that
\bea\label{yuk}
Y^{~~j}_{U~i} &=& \eta_U \, g^{~~j}_{U~i} + \eta'_U \, g^{~~j}_{U~k} [(g_U^\dagger)^{k}_{~l} \, \, (g_U)^{l}_{~i}] + \cdots, \nn \\
Y^{~~j}_{D~i} &=& \eta_D \, g^{~~j}_{D~i} + \eta'_D \, g^{~~j}_{D~k} [(g_U^\dagger)^{k}_{~l} \, \, (g_U)^{l}_{~i}] + \cdots.
\eea
We decompose the second scalar doublet as 
\bea
S = \left(\begin{array}{c} {S^+} \\
{S^0} \end{array} \right), 
\eea
where $S^{0} = (S^{0} _R + i S^{0}_I)/\sqrt{2}$. 

The scalar potential is
\bea\label{pot}
V &=& \frac{\lambda}{4} \, \left(H^{\dagger \, i} \, H_i - \frac{v^2}{2} \right)^2 + m_1^2 \, (S^{\dagger i} \, S_i),  \\
&+& (m_2^2 \, H^{\dagger \, i} S_i + {\rm h.c.}) + \lambda_1 \, (H^{\dagger \, i} H_i) \, (S^{\dagger \, j} S_j), \nn \\
&+& \lambda_2 \, (H^{\dagger i} \, H_j) \, (S^{\dagger j} \, S_i)  + \left[\lambda_3 H^{\dagger i} \, H^{\dagger j} \, S_i \, S_j + {\rm h.c.} \right], \nn \\
&+& \left[\lambda_4 H^{\dagger i} \, S^{\dagger j} \, S_i \, S_j + \lambda_5 S^{\dagger i} \, H^{\dagger j} \, H_i \, H_j + {\rm h.c.} \right], \nn \\
&+& \lambda_6 (S^{\dagger i} S_i)^2.\nn
\eea
where $i,j$ are $\rm SU(2)$ indices.  Here $v \simeq 246{\rm GeV}$ is the vacuum expectation value (vev) of the Higgs. Since we adopted the convention that the doublet $S$ does not get a vev the parameters $m_2^2$ and $\lambda_5$ are related by,
\begin{equation}
m_2^2+\lambda_5^\star{v^2 \over 2}=0.
\end{equation}
The spectrum of neutral real scalars consists of the Higgs scalar $h$  and $S_R^0$ and $S_I^0$. However, these are not mass eigenstates. In the $(h, S_R^0,S_I^0)$ basis, the neutral mass squared matrix ${\cal M}^2 $, where $\lambda_3$ is chosen real and positive, is

\begin{equation}
{\cal M}^2 =\begin{pmatrix}
		m_h^2  &  \lambda_5^Rv^2 & \lambda_5^Iv^2   \\
		  \lambda_5^Rv^2 &m_S^2+\lambda_3 v^2 & 0  \\
		\lambda_5^Iv^2  & 0 & m_S^2-\lambda_3^2 \\
	\end{pmatrix}.
	\end{equation}
Where $m_S^2 = m_1^2 + (\lambda_1 + \lambda_2) v^2/2$.	
Within the convention that $\lambda_3$ is real, the  couplings  $\eta_U$, $\eta'_U$, $\eta_D$ and $\eta'_D$ and $\lambda_5=\lambda_R^R + i \lambda_5^I$  are in general complex. 
The mass eigenstate scalars $N_j^0$ with mass $m_j$ are related to $h, S_R^0, S^0_I$ by 
the orthogonal transformations
\bea
h &=& \sum_j O_{hj} \, N_j, \nn \\
S_R &=& \sum_j O_{Rj} \, N_j,  \quad \quad S_I = \sum_j O_{Ij} \, N_j.
\eea
We find the  CP violating  NP contribution to $B_q-{\bar B}_q$ mixing from neutral scalar exchange is
\bea\label{o2coeff}
C^{\rm NP}(m_t)= \left(\sqrt{2} \,  \eta'_D \, m_b/v \,\right)^2  \left( F \left( \sqrt{2} m_t/v\right)\right)^2  \frac{\Delta}{2},
\eea
where $ F(x)=x^2+ ~...~,$ and
\bea
\Delta = \sum_j \frac{(O_{Rj} + i \, O_{Ij})^2}{m_j^2}.
\eea
For the rest of  this paper we truncate the expansions in ${\sqrt 2} m_t/v$ at the leading non trivial term. So, for example, in Eq.~(\ref{o2coefff}) we use $F(x)=x^2$.

For simplicity we now focus on the case where $\lambda_5=0$. Then $h$, $S_R^0$  and $S_I^0$ are mass eigenstates and the term in the potential proportional to $\lambda_3$ is of interest as it leads to the mass splitting between the real neutral fields 
given by $m^2_R - m^2_I = 2\lambda_3 \, v^2$. When this term is non vanishing\footnote{Note that imposing custodial symmetry on the potential does not force $\lambda_3 \rightarrow 0$. Custodial symmetry violation is a measure of the total mass splitting $(m_R^2 - m^2_\pm) (m^2_I - m_\pm^2) \propto (\lambda_2^2 - (2 \lambda_3)^2) \, v^4$ in terms of the potential given in Eq. (\ref{pot}).}, the tree level
exchange of $S_{R/I}$ generates a CP violating  NP contribution to $B_q-{\bar B}_q$ mixing and
\bea\label{o2coefff}
C^{\rm NP}(m_t) =(\eta'_D)^2  \left(\frac{\sqrt{2} \, m_t}{v}\right)^4  \left(\frac{\lambda_3 \, m_b^2}{m_S^4 - \lambda_3^2 \, v^4}\right).
\eea
In the above equation, the bottom quark mass $m_b\simeq 2.93~{\rm GeV}$ is evaluated at the top quark mass scale and $m^2_{S_{R/I}}=m_S^2 \pm \lambda_3 v^2$.

Using Eq.~(\ref{key}) the mass scale of the new scalars is given by
\bea
m_S^4  \simeq  {20\, \pi^2\, \lambda_3 \, |\eta'_D|^2 \eta' m_b^2 \, m_t^4 \over h_q \, \eta \,M_W^2 \, S(m_t^2/m_W^2)}  + \lambda_3^2 v^4.
\eea
Using the best fit value $h_q =0.255$ \cite{Lenz:2010gu} we find that
\bea
\label{upperlimitmass}
m_S^4 \simeq  (154\, {\rm GeV})^4  \,|\eta'_D|^2 \lambda_3+ (246 \, {\rm GeV})^4 \, \lambda_3^2.
\eea
Then for example with a value  $|\eta'_D| = 5$  and $\lambda_3=1$ the scalar mass scale is $m_S \simeq 360 \,  {\rm GeV}$.
As the mass splitting is significant we show in Fig. 1 the masses of the neutral scalars $S_R,S_I$ as a function of $\eta_D'$.
Moderate enhancements of $\eta_D'$ avoid a light neutral state.
\begin{figure}[hbtp]
\centerline{\scalebox{1}{\includegraphics{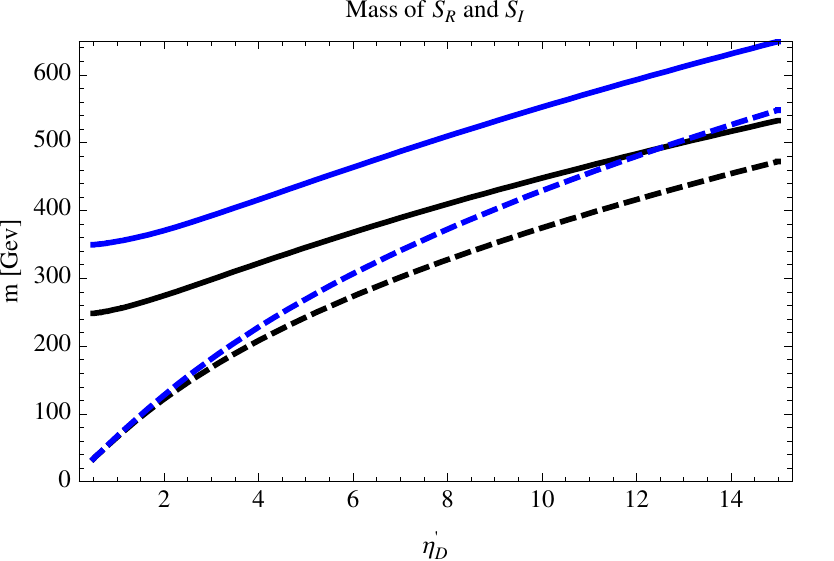}}}
\caption{Mass of $S_R$ (solid) and $S_I$ (dashed) as a function of $\eta_D'$ for fixed $\lambda_3$.  The upper (blue curves) are for $\lambda_3 = 1$
the lower (black) curves are for $\lambda_3 = 0.5$.}
\end{figure}

We have checked that the mass scale $m_S$ required for  $B_q-{\rm B}_q$ mixing is compatible with the constraints from $K - \bar{K}$ mixing.
This compatibility is due to MFV, which causes the ratio of the relevant Wilson coefficients to scale as $m_s^2/m_b^2 \, |V_{td}/V_{tb}|^2$.

Next  we derive constraints on $\eta_U \eta_D$ that come from one loop Feynman diagrams with charged $S$ scalar exchange. We will show that limits on electric dipole moments  imply that 
$|{\rm Im}[\eta_U \eta_D]|\lesssim 10^{-1}$.  Note that writing this as a constraint just on $\eta_U \eta_D$  depends on  truncating a function of $\sqrt{ 2}m_t/v$ at leading order. We also examine the constraint on $ {\rm Re}[\eta_U \eta_D]$  coming from experimental data on weak radiative $B$ decay.

\subsubsection{Neutron Electric Dipole Moment}

The large CP violating phases needed in this two scalar doublet model contribute to other CP violating observables. Notable among them are electric dipole moments  (EDM's).  We will restrict our discussion here to the dominant contribution that is not suppressed by small quark masses when naive dimensional analysis (NDA) \cite{Manohar:1983md} is used. It comes about through the colour electric dipole moment of 
the b quark  \cite{Boyd:1990bx,Braaten:1990gq} due to the effective Hamiltonian
\bea
\delta {\mathcal{H}}_{bg} = C_{gb} \, g_3 \, m_b \, \bar{b} \, \sigma_{\mu \, \nu} \, T_a \, G^a_{\lambda \, \sigma} \, \epsilon^{\mu \, \nu \, \lambda \, \sigma} \, b,
\eea
inducing the dimension six CP violating operator
\bea
O_G =   g_3^3 \, f_{\alpha \, \beta \, \gamma} \, \epsilon^{\mu \, \nu \, \lambda \, \sigma} \, G_{\alpha \, \mu \, \rho} \, G_{\beta \, \nu}^{\rho} \, G_{\gamma \, \lambda \, \sigma},
\eea
of Weinberg \cite{Weinberg:1989dx} when the $b$ quark is integrated out. Our discussion will largely parallel
the discussion of  \cite{Manohar:2006ga}. As $\rm S$ couples both to the up and down type quarks it
induces a one loop contribution to the effective Hamiltonian above with
\bea
C_{gb}(m_S) = \frac{{ -{\rm Im}[ \eta_U^\star \, \eta_D^\star]}}{64 \pi^2 \, m_{S^\pm}^2}  \, \left(\frac{\sqrt{2} \, m_t}{v} \right)^2 \, f(m_t^2/m_{S^\pm}^2),
\eea
where
\bea
f(x) = \frac{\log x}{(x -1)^3}  + \frac{x - 3}{2 \, (x -1)^2}.
\eea
Running to $\mu \sim m_b$ using \cite{Boyd:1990bx, Braaten:1990gq} and estimating the 
matrix element of the operator with NDA\footnote{We use method (a) of \cite{Manohar:2006ga} with $\alpha_s(\mu = 1 \, {\rm GeV}) \sim 4 \, \pi$.} gives in $\rm e$-$\rm cm$ units
\bea
\small
d_n \sim 2 \, {\rm Im}[\eta_U^\star \, \eta_D^\star] \, f(m_t^2/m_{S^\pm}^2) \, \left(\frac{1 \, {\rm TeV}}{m_{S^\pm}} \right)^2  \, 10^{-26}.
\eea

This is a significantly larger effect on EDM's than quoted in the general operator analysis \cite{Blum:2010mj}
examining the effects of four Fermi operators on EDM's as this contribution is not
suppressed by small mixing angles or light quark masses.
For $m_{S^\pm}=360~{\rm GeV} $ the neutron EDM experimental bound of $d_n < 2.9 \times 10^{-26} \, {\rm e}$-$\rm cm$ implies that $|{\rm Im}[ \eta_U^* \eta_D^*]|<0.26$  
We plot the allowed $|{\rm Im}[ \eta_U^* \eta_D^*]|$ as a function of mass for this NDA estimate in Fig. 2.
\begin{figure}[hbtp]
\centerline{\scalebox{1}{\includegraphics{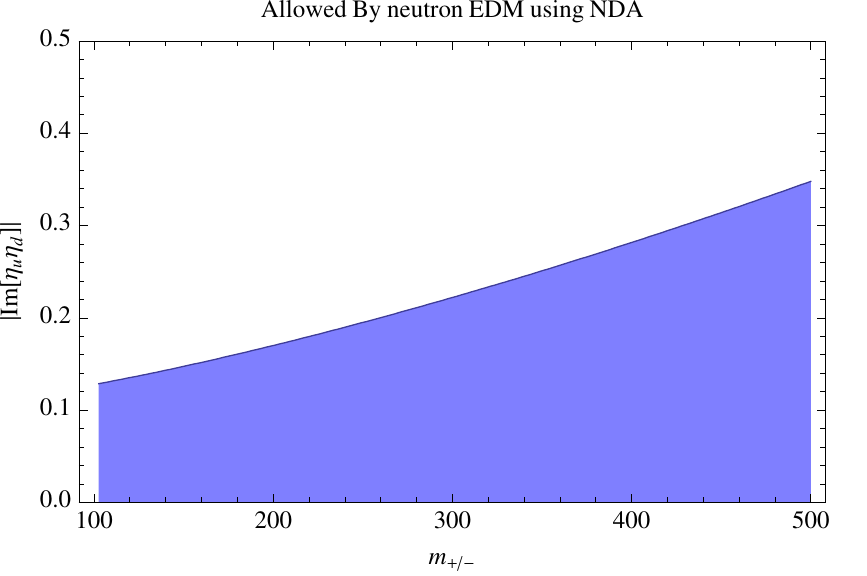}}}
\caption{Allowed $|{\rm Im}[ \eta_U^* \eta_D^*]|$ as a function of the charged scalar mass.}
\end{figure}

This suggests that $|{\rm Im}[\eta'^*_U \eta'^*_D]|$ (and the sum of the effect of all other cross terms such as $|{\rm Im}[\eta''^*_U \eta''^*_D]|$ etc.) is also small.  However, given the uncertainties from hadronic  matrix elements and given the fact that the parameters that enter the  contribution to EDM's are not identical to those in $B_q-{\bar B}_q$ mixing  it is difficult to draw precise conclusions on the parameters in the model that are important for mixing.


\subsubsection{$B \rightarrow X_s \, \gamma$ Constraints}

Of course the two scalar doublet model also gives new contributions to quantities that are not $\rm CP$ violating.  Here we briefly review the constraints on this model from $B \rightarrow X_s \, \gamma$
with these assumptions. The extra term in the effective Hamiltonian arises from charged scalar exchange and has the form,
\bea
\delta \, {\mathcal{H}}_{{\bar{B}}\rightarrow X_s \, \gamma} =  [V_{ts}^\star \, V_{tb}] C_\gamma \, \left(\frac{e \, m_b}{16 \, \pi^2} \, \bar{s}_L \, \sigma_{\mu \, \nu} \, F^{\mu \, \nu} \, b_R \right), 
\eea
where $e < 0$ is the electric charge. The Wilson coefficient is given by
\bea
C_\gamma = \eta_U^\star \, \eta_D^\star \,  \left(\frac{2 \, m_t^2}{v^2} \right) \, \frac{f_\gamma(m_t^2/m_{S^\pm}^2)}{3 \, m_{S^\pm}^2},
\eea
with
\bea
\small
f_\gamma(x) = \frac{1}{4} \left(\frac{1 + 2 \, x \, {\rm log}x - x^2}{(1-x)^3} \right) - \left(\frac{1 +{\rm log} x - x}{(1-x)^2} \right).
\eea
This operator's contribution 
to the measured branching fraction ${\rm BR}(\bar{B} \rightarrow X_s \, \gamma)_{E_\gamma > 1.6 \, {\rm GeV}}$ is known \cite{Grzadkowski:2008mf}
\bea
\frac{{\rm BR}(\bar{B} \rightarrow X_s \, \gamma)_{E_\gamma > 1.6 \, {\rm GeV}}}{10^{-4}} = 3.15 \pm 0.23 - 4.0 v^2 \, C_\gamma. \nn
\eea
This constraint includes the effect of running this operator down to the scale $m_b$. Comparing to the
world experimental average \cite{Barberio:2007cr}  we obtain a $1 \sigma$ bound on the parameters of the form
\bea
 -0.17 < Re[\eta_U^\star \, \eta_D^\star] \, f_\gamma(m_t^2/m_{S^\pm}^2)  \, \frac{m_t^2}{3 \, m_{S^\pm}^2} < 0.07.
\eea
For $m_{S^\pm} = 360 ~{\rm GeV}$ we find $-1.7 < {\rm Re}[\eta_U^\star \eta_D^\star]   < 0.7$. Although this constraint is weak it is interesting that EDM's constrain ${\rm Im}[\eta_U^\star \, \eta_D^\star]$ while $B \rightarrow X_s \, \gamma$
constrains ${\rm Re}[\eta_U^\star \, \eta_D^\star]$.

\subsubsection{Collider Physics: Two Scalar Doublet Model}

Pairs of $S$ particles can be produced through the tree level exchange
of vector bosons  produced through $q \, \bar{q}$ initial states in the case of the Tevatron
and LHC and $e^+ \, e^-$ in the case of LEPII.  

From LEPII a bound on the mass scale of the new scalar doublet is obtained as no anomalous two and four jet events were seen
when operating at $\sqrt{s} = 209 \, {\rm GeV}$ where $0.1 \, fb^{-1}$ of integrated luminosity was collected.
The relevant cross sections in this case are given in \cite{Burgess:2009wm} and the masses are bound to 
be 
\bea
m_{S^\pm} \gtrsim 105 \, {\rm GeV}, \quad m_{S_R^0} + m_{S_I^0} \gtrsim 209 \, {\rm GeV}. 
\eea
We plot the allowed $m_S,\lambda_3$ that satisfy this second bound for the minimal two scalar doublet model in Fig.3.
We have also performed an electroweak precision data fit. For scalar masses $\sim 100 \,{\rm GeV}$
the constraints are weak. The allowed mass splitting 
in this model is $|m_I - m_\pm| \lesssim 200 \, {\rm GeV}$ using the $95 \% {\rm CL}$ region.
\begin{figure}[hbtp]
\centerline{\scalebox{1}{\includegraphics{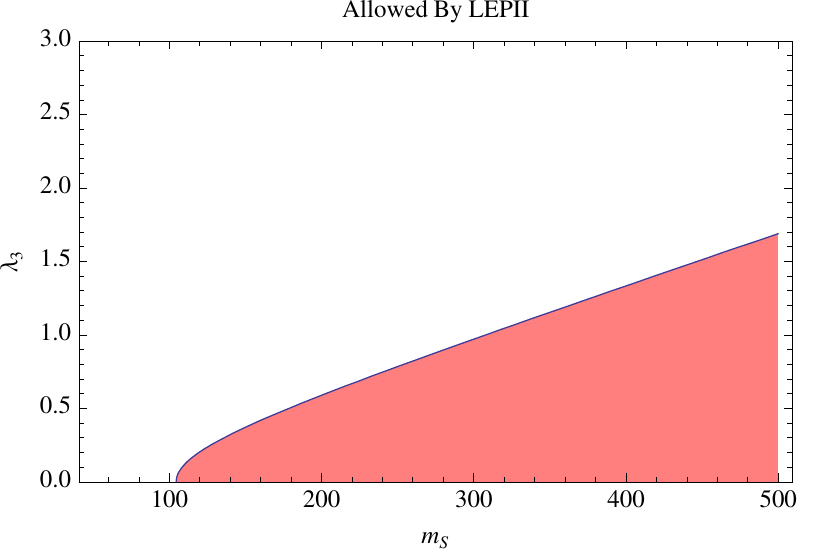}}}
\caption{The allowed range of parameters $m_S,\lambda_3$ in the two scalar doublet model considering LEPII direct production bounds.}
\end{figure}

A light mass of $S^0_I$ is allowed as these states must be produced in pairs through vector boson exchange and $S^0_R$ can be heavy.
However, a single neutral scalar particle can be produced at 
the Tevatron in association with a charged scalar though $W^\pm$ exchange. The partonic production cross section for producing $S^\pm \, S^0_{I}$ or 
$S^\pm \, S^0_{R}$ (when the width is neglected) is
\bea\label{wcross}
\sigma = \frac{(p^2/s)^{3/2}}{s_W^4} \, \left(\frac{\pi \, \alpha_{e}^2(M_Z)}{6 \, s}\right)\, \bigg\vert 1-\frac{M_W^2}{s} \bigg\vert^{-2},
\eea
where $p$ is the center of mass momentum of one of the produced particles and $s$ is the partonic center of mass energy squared.  We scan over the
parameter space allowed by LEPII using this formula for the Tevatron production cross section (with MSTW 2008 PDF's  \cite{Martin:2009iq}) where the $W^\pm$ is produced off the valence $u,d$ quarks.
The renormalization scale in what follows is always varied between $m_S/2$ and $2 \, m_S$. The cross sections as a function of mass are shown in Fig. 4.
\begin{figure}[hbtp]
\centerline{\scalebox{1}{\includegraphics{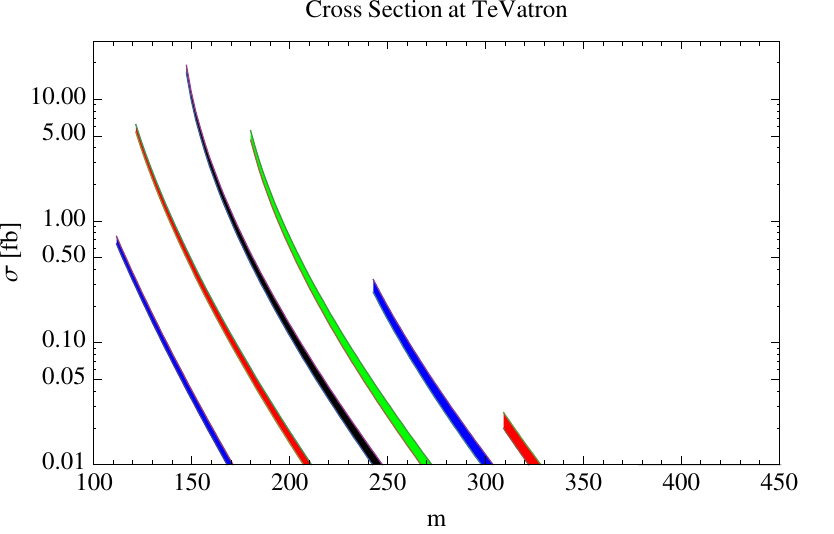}}}
\caption{The cross section $\sigma(S^\pm \, S_I) + \sigma(S^\pm \, S_R)$ as a function of $m_S$ for $\lambda_3 = \( 0.1,0.2,0.35,0.5,0.75,1\)$ going left to right. Here we have also imposed custodial symmetry on 
the potential $\lambda_2 = \pm 2 \lambda_3$  for simplicity in the parameter scans.}
\end{figure}
The Tevatron can potentially constraint some of the allowed parameter space. 
Search strategies for pair production through weak boson fusion of charged $S^\pm$ particles that decay into $\bar{t} \, b \, \bar{b} \, t$
are also somewhat promising. In this case the production cross section for $m_S \sim 200 \, {\rm GeV}$ is $\sigma \sim 1 \, {\rm fb} $ with a signal
of two $b$ jets and two $t$ jets is produced in association with tagging light quark jets at large $\rm p_T$.

At the LHC, production through the tree level exchange of a vector boson is no longer dominated by $W^\pm$ exchange. The cross sections for the pair production 
of scalars are all similar in their dependence on $\lambda_3$ and as a function of $m_S$. We show $\sigma(pp \rightarrow W^\pm \rightarrow S^\pm \, S_{R/I})$ 
for $\sqrt{s} = 7 \, {\rm TeV}$ in Fig. 5. 

\begin{figure}[hbtp]
\centerline{\scalebox{1}{\includegraphics{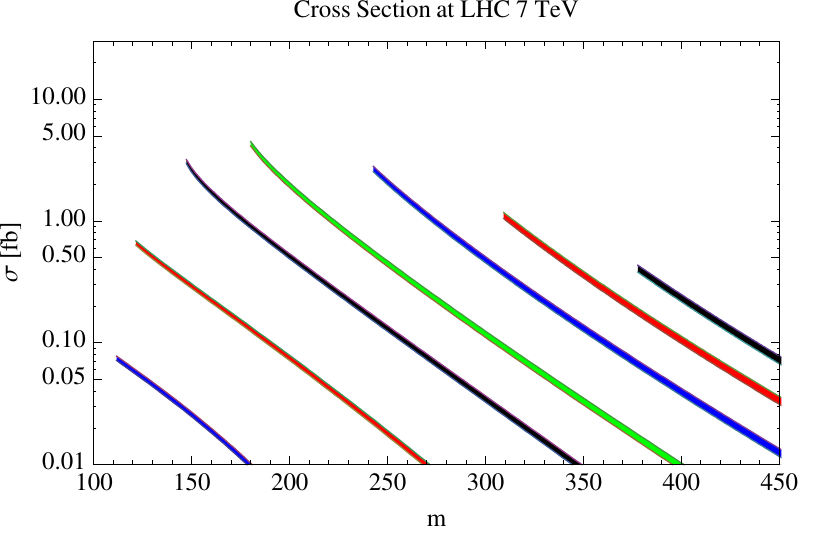}}}
\caption{The cross section $\sigma(S^\pm \, S_I) + \sigma(S^\pm \, S_R)$ as a function of $m_S$ for $\lambda_3 = \( 0.1,0.2,0.35,0.5,0.75,1,1.25\)$ going left to right. Here we have also imposed custodial symmetry on 
the potential $\lambda_2 = \pm 2 \lambda_3$  for simplicity in the parameter scans. The other production cross sections through $Z^\star,\gamma^\star$ are similar.}
\end{figure}

Although these vector boson exchange cross sections for LHC are small, potentially observable signals at LHC do exist  when $\eta_D$ is larger than one, and $S_{R/I}$ is made with
large logarithms associated with collinear gluon splitting \cite{Dawson:2005vi} and small $p_T$ of the spectator $b$ quarks. The cross-section for the production of the lightest state
$b\, \bar{b}\to S_I^0$ at leading log takes the form \cite{Mantry:2007ar}
\begin{eqnarray}
\label{bbS}
\sigma(b \bar{b} S_I^0) \simeq \frac{|\eta_D|^2 \pi} {3 \, s} \left(\frac{m_b^2}{v^2}\right) \int _{\frac{m_{S_I}^2}{s}}^1 \frac{dx}{x} b(x,\mu) \bar{b}(\frac{m_{S_I^0}^2}{x s},\mu),
\end{eqnarray}
where $b(x,\mu)$ and $\bar{b}(x,\mu)$ are the
$b$ quark and antiquark PDFs respectively. 
The large logs from collinear gluon splitting are summed into the parton distribution functions by choosing $\mu \sim m_S$.
When we let $\eta_D = \sqrt{m_S/(154 \, {\rm GeV})}$ and choose $\lambda_3 =1$ the production cross sections for the
LHC are given by Fig.6. This production mechanism must compete with the large $b$ production background from QCD. However, we note that this signal has a distinct feature in its reconstruction of 
a resonance in the highest $p_T$ $b$ quark pair with a larger percentage of its total number of events
at high $p_T$ and small rapidity than the SM background, which has an approximate Rutherford scattering angular dependence in its production of $b$ quarks.
\begin{figure}[hbtp]
\centerline{\scalebox{0.85}{\includegraphics{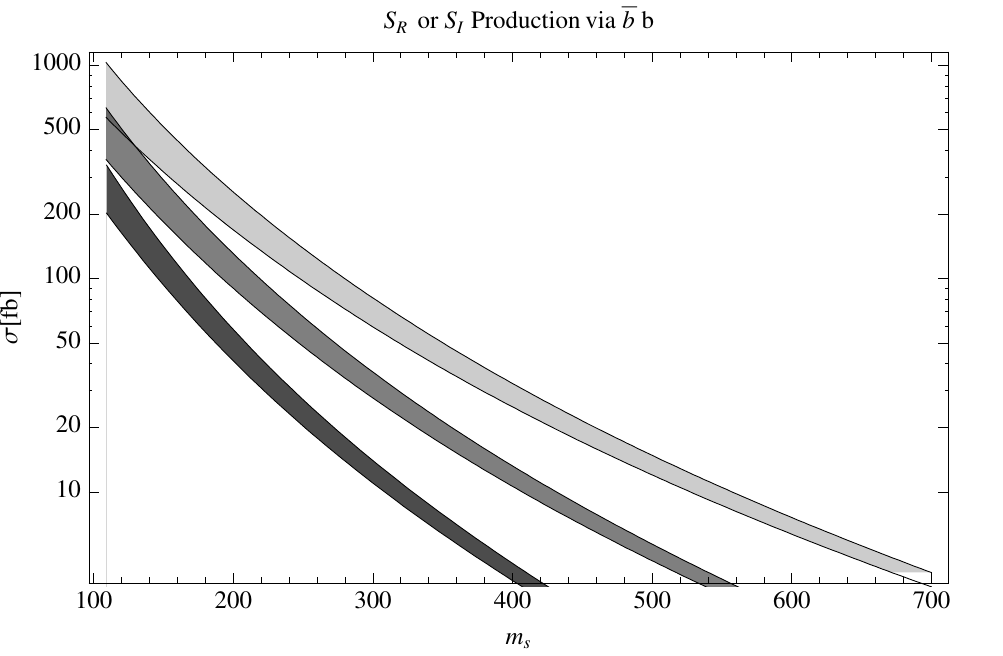}}}
\caption{The cross section $\sigma(p \, p \rightarrow b \, \bar{b} \, S_I^0)$ for $S_I^0$ produced through collinear gluon splitting and b quark fusion. Shown are the cross sections
for $\sqrt{s} = 7,10,14 \,{\rm TeV}$.}
\end{figure}

\section{A Model  with $\eta_U$ Naturally Small}

The charged scalar in the two Higgs doublet model has couplings to the quarks that (at leading order in the Yukawa matrices) are given by,
\begin{equation}
{\cal L}_{\rm charged}=\eta_U {\bar u}_R  g_U  d_L  S^+ + \eta_D{\bar d}_R g_D u_L S^- +{\rm h.c.}
\end{equation}
We need large CP violating phases to  get the fit value of $\sigma_q$. For large phases the limits on electric dipole moments  suggest that
$|\eta_D \eta_U| \lesssim 10^{-1}$ for charged scalars with mass of a few hundred ${\rm GeV}$.  Unless the charged scalars are considerably heavier than the neutral ones  this bound is expected to hold in the minimal two scalar doublet model if the model is to give the central value for $h_q$.  If the limits on the electric dipole moments improve then this may become a more serious constraint. 

There does not seem to be any acceptable symmetry reason that this product is small and at the same time $\eta_D$ is enhanced. 
In order to see that this is the case it is convenient to work in the basis where both $H$ and $S$ get a vevs $v_H$ and $v_S$ respectively. These can be chosen to be real. Then the charged scalar $P^+$ is the linear combination of the fields $h^+$ and $S^+$,
\begin{equation}
P^+={v_Sh^+-v_H S^+ \over \sqrt{v_H^2+v_S^2}}
\end{equation}
Because $H$ no longer plays a special role we write the couplings  of the scalars to the quarks as,
\bea
\label{general}
{\cal L}_Y&=& \epsilon_H{\bar u} {\tilde g}_{U}Q_L H+ \epsilon_S{\bar u}_{R} {\tilde g}_{U}  Q_{L } S   \\
&+&  \epsilon'_H {\bar d}_{R}  {\tilde g}_DQ_{L} H^{\dagger}+  \epsilon'_S {\bar d}_{R}  {\tilde g}_D Q_LS^{\dagger}+{\rm h.c.} \nn
\eea
Here we are using  MFV and taking the quantities that break the flavor symmetry to be ${\tilde g}_{U/D}$ . These matrices are proportional to the usual Yukawa matrices,
\bea
g_U \! \! = \! \! \left({ \epsilon_H v_H+\epsilon _S v_S} \over {\sqrt{v_H^2+v_S^2}}\right){\tilde g}_U, 
g_D \! \! = \! \! \left({ \epsilon'_H v_H+\epsilon' _S v_S} \over {\sqrt{v_H^2+v_S^2}}\right){\tilde g}_D. 
\eea
Writing the charged scalar interaction as
\begin{equation}
{\cal L}_{\rm charged}=\eta_U{\bar u}_R  g_U  d_L  P^+ + \eta_D{\bar d}_R g_D u_L P^- +{\rm h.c.}
\end{equation}
we find that,
\bea
\eta_U={\epsilon_Hv_S- \epsilon_S v_H \over \epsilon_Hv_H+\epsilon_S v_S}, \quad \eta_D={\epsilon'_Hv_S- \epsilon'_S v_H \over \epsilon'_Hv_H+\epsilon'_S v_S}.
\eea
One way to get $\eta_U$ small while $\eta_D$ is large is to have $v_H \gg v_S$ so that $\eta_U \sim  -\epsilon_S/\epsilon_H$ and $\eta_D \sim \epsilon'_H/\epsilon'_S$ and take the corresponding ratios of $\epsilon$'s to be small and large respectively.   For their product to be small we also need,   $\epsilon_S \epsilon_H'/\epsilon_S' \epsilon_H$ to be small.   This is clearly possible,  however there doesn't appear to be any symmetry  reason behind these choices.

Note that there is an interchange symmetry where $H \leftrightarrow S$ that forces both $\eta_U=\eta_D=0$. But  in the limit of that symmetry, $P^+=(h^+-S^+)/\sqrt{2}$, and the symmetry's action on $P^+$ is  $P^+ \rightarrow -P^+$. Hence  there is a stable charged scalar.

The Glashow-Weinberg model \cite{Glashow:1976nt}  where $H$ couples to the $u$-type quarks and $S$ couples to the $d$-type quarks has, $\epsilon'_H=\epsilon_S=0$ and so  $\eta_U=v_S/v_H$ and $\eta_D=-v_H/v_S$. In this model $\eta_D \eta_U=-1$, so when $\eta_U$ is small  $\eta_D$ is large, but their product cannot be made small even with a tuning of parameters.

In this section we construct a $\rm MFV$ model  that has $\eta_U$ small for a symmetry reason.  New scalars that transform under 
flavour \cite{Arnold:2009ay} can naturally have a small $\eta_U$. Consider a scalar field $S_8$ that transforms the same way as the Higgs doublet under the gauge group but as $({\bf 1},{\bf 8},{\bf 1})$ under the flavor group,
\bea
S_8 \rightarrow V_D \, S_8 \, V_D^\dagger.
\eea
We choose to represent the scalar in terms of the Gell-Mann matrices $S_8 = S_8^a \, T^a$ where $a = 1, ...,8$ is a flavour index.
The Yukawa couplings are given by
\bea\label{general}
{\cal L}_Y &=&  {\bar u}_{R}^i \, \hat{Y}^{~l}_{U~i} \, (g_D^\dagger)^o_{~l}   \, (T^{a})^n_{~o} \, (g_D)^j_{~n}   \, Q_{Lj}  \, S_8^a, \\
&\,&  + {\bar d}_{R}^i \, (T^a)^m_{~i} \, (\hat{Y}_D)^j_{~m}  Q_{Lj}  \, S^{\dagger \, a}_8+{\rm h.c.} \nn
\eea
where we have made the flavour indices explicit. We use hat superscripts to distinguish this model's parameters from the 
two scalar doublet model.  Recall that,
\bea\label{yuk1}
{\hat Y}^{~~j}_{U~i} &=& {\hat \eta}_U \, g^{~~j}_{U~i} + {\hat \eta}'_U \, g^{~~j}_{U~k} [(g_U^\dagger)^{k}_{~l} \, \, (g_U)^{l}_{~i}] + \cdots, \nn \\
{\hat Y}^{~~j}_{D~i} &=& {\hat \eta}_D \, g^{~~j}_{D~i} + {\hat \eta}'_D \, g^{~~j}_{D~k} [(g_U^\dagger)^{k}_{~l} \, \, (g_U)^{l}_{~i}] + \cdots.
\eea

The potential is given by
\bea
V &=& \frac{\lambda}{4} \, \left(H^{\dagger i} \, H_i - \frac{v^2}{2}\right)^2 + 2 \hat{m}_1^2 \, {\rm Tr}[S^{\dagger \, i}_8 \, S_{8 \, i}] \nn \\
&+&  \hat{\lambda}_1 \, H^{\dagger i} \, H_i \, {\rm Tr}[S^{\dagger j}_8 \, S_{8 \,j}] +  \hat{\lambda}_2 \, H^{\dagger i} \, H_j \, {\rm Tr}[S^{\dagger j}_8 \, S_{8 \,i}] \nn \\ 
&+&   \left[\hat{\lambda}_3 \, H^{\dagger i} \, H^{\dagger j} \, {\rm Tr}[S^{\dagger}_{8 i} \, S_{8 \,j}] +  \hat{\lambda}_4 \, H^{\dagger i} {\rm Tr}[S^{\dagger \, j}_{8}  S_{8 \,j} S_{8 \,i}] \right. \nn \\ 
&+&   \left. \hat{\lambda}_5 \, H^{\dagger i} \, {\rm Tr}[S^{\dagger j}_{8}  S_{8 \,i} S_{8 \,j}]  + {\rm h.c.} \right]   \\ 
&+&   \hat{\lambda}_6 \, {\rm Tr}[S^{\dagger i}_8 \, S_{8 \,i} \, S^{\dagger j}_8 \, S_{8 \,j}] +  \hat{\lambda}_7 \, {\rm Tr}[S^{\dagger i}_8 \, S_{8 \,j} \, S^{\dagger j}_8 \, S_{8 \,i}]  \nn \\ 
&+&   \hat{\lambda}_8  {\rm Tr}[S^{\dagger i}_8  S_{8 \,i}] \, {\rm Tr}[S^{\dagger j}_8 S_{8 \,j}] +   \hat{\lambda}_9 {\rm Tr}[S^{\dagger i}_8 S_{8 \,j}] {\rm Tr}[S^{\dagger j}_8 S_{8 \,i}] \nn \\
&+&   \hat{\lambda}_{10} {\rm Tr}[S_{8 i} S_{8 \,j}] \, {\rm Tr}[S^{\dagger i}_8 S^{\dagger j}_8] + \hat{\lambda}_{11} {\rm Tr}[S_{8 i}  S_{8 \,j}]  {\rm Tr}[S^{\dagger j}_8 S^{\dagger i}_8]. \nn
\eea
In the potential the index is an $\rm SU(2)$ index and the trace is over the down flavour index.  We again  rotate the phase of $S_8$ (relative to $H$) so that the $\hat{\lambda}_3$ term is real, then
the  couplings and $\hat{\lambda}_{4,5}$ and the $\eta$'s are in general complex.  In the above potential there are no linear terms in $S_8$ after $H$ gets its vacuum expectation value and so it is natural for it not to have a vev.

In the potential and the ${\hat Y}'s$  one can also insert arbitrary numbers of $g_D \, g_D^\dagger$ matrices between contractions of a down index. We work in the down basis
so that $g_D = {\rm diag}(\sqrt{ 2}m_d/v,\sqrt{2}m_s/v,\sqrt{2}m_b/v)$. The interactions in the potential do not change flavour and  
are suppressed by $m_b^2/v^2$ so we neglect them. 

Keeping just the leading term in, $\sqrt{2}m_t/v$, the Wilson coefficient of the effective Hamiltonian as defined in Eq. (\ref{NPH}) is
\bea\label{o2coeff}
C^{\rm NP}(m_t) =  \frac{ (\hat{\eta}'_D)^2\left(\sqrt{2} \, m_t/v\right)^4 \, \hat{\lambda}_3 \, m_b^2/6}{\hat{m}_S^4 - \hat{\lambda}_3^2 \, v^4/4}.
\eea
where $\hat{m}_S^2 = \hat{m}_1^2 + \left(\hat{\lambda}_1 + \hat{\lambda}_2\right) v^2/4$.
This leads to the mass bound
\bea\label{upperlimitmass}
\hat{m}_S^2 &\simeq&  (98 \, {\rm GeV})^4  \,|\hat{\eta}'_D|^2 \hat{\lambda}_3 + \, (174 \, {\rm GeV})^4 \, \hat{\lambda}^2_3.
\eea
in terms of the parameters defined in the potential. The mass spectrum of 
the new doublet is given by
\bea
m_{S^\pm}^2 &=& \hat{m}_S^2 - \hat{\lambda}_2 \, \frac{v^2}{4}, \nn \\
m_{S_R^0}^2 &=& \hat{m}_S^2 + \hat{\lambda}_3  \frac{v^2}{2}, \nn \\
m_{S_I^0}^2 &=& \hat{m}_S^2 -  \hat{\lambda}_3 \frac{v^2}{2}, 
\eea
We show the masses of the neutral scalars for this model in Fig.7.
\begin{figure}[hbtp]
\centerline{\scalebox{1}{\includegraphics{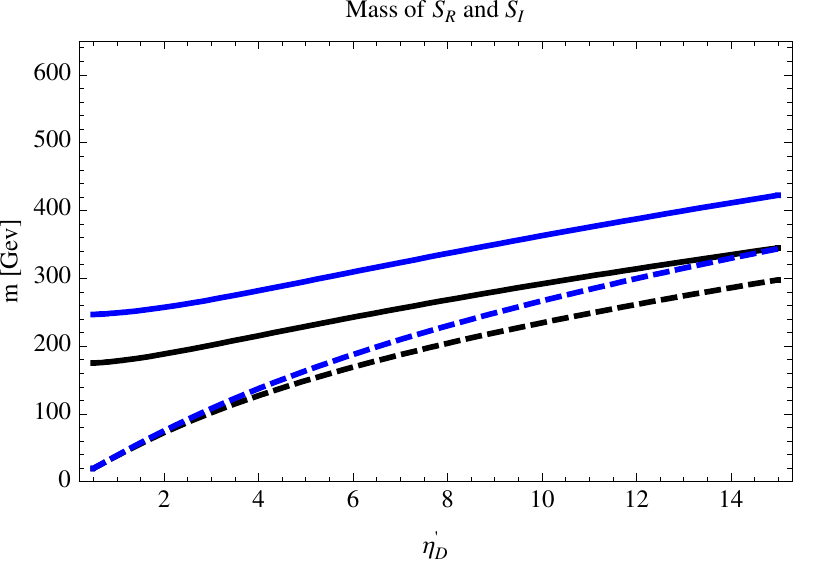}}}
\caption{Mass of $S_R$ (solid) and $S_I$ (dashed) as a function of $\eta_D'$ for fixed $\lambda_3$.  The upper (blue curves) are for $\lambda_3 = 1$
the lower (black) curves are for $\lambda_3 = 0.5$.}
\end{figure}

This model has eight new scalar doublets. Nevertheless, precision electroweak constraints are satisfied (when the Higgs is fixed to be $m_h = 96^{+29}_{-24} \, {\rm GeV}$ for $m_S \gtrsim 100 \, {\rm GeV}$)  when $|m_I - m_\pm| < 50 \, {\rm GeV}$ \cite{Burgess:2009wm}.
Conversely, custodial $\rm SU(2)$ violation in such a light scalar doublet leading to a positive contribution to $\Delta T$ can raise the allowed
mass of the Higgs in EWPD \cite{Peskin:2001rw,Burgess:2009wm}.

In this model the coupling constant analogous to  $\eta_U$ is naturally of order $(m_b/v)^2 \sim 10^{-3}$ and
the $\bar{B} \rightarrow X_S \, \gamma$ and neutron EDM  effects of the model
are suppressed as phenomenologically required due to MFV. 

The collider phenomenology in this model is very similar to the discussion on the two scalar doublet  model.
The LEPII constraints allow a larger parameter space due to the smaller mass splitting.
The main differences for the Tevatron is that the cross sections we have discussed are increased by an order of magnitude
due to the larger flavour representation. Slightly smaller production cross sections through $b$ quark fusion with low $p_T$ spectator $b$ quarks
are expected at LHC as the normalization of the Gell Mann matrix decreases the cross section by a factor of three.

We consider speculation on the UV origin of such a $S_8$ doublet, or other accompanying non flavour singlet doublets that transform under the $\rm SU(3)_{U_R}$ as
an $\bf 8$ to be premature and beyond the scope of this work.

\section{Conclusions}
In this paper we discussed the new (i.e., beyond the minimal standard model) physics in the region of parameter space for which the two scalar doublet model with MFV  gives the additional contributions  to $B_q- {\bar B}_q$ mixing  that are hinted at by the data on flavor physics in the $B$-sector. It requires additional light scalars that may be discovered at the Tevatron or LHC. Experimental limits on electric dipole moments suggest a region of parameter space that can occur naturally in some models where the new doublet of scalars transforms non-trivially under the flavour group. We constructed such a model.

\section*{Acknowledgements}
We thank Z. Ligeti and Jim Cline for comments on the manuscript.
This work was partially supported by funds from the
Natural Sciences and Engineering Research Council (NSERC) of
Canada. Research at the Perimeter Institute is supported by the
Government of Canada through Industry Canada and by the Province
of Ontario through the Ministry of Research \& Innovation.  MBW is greatful to Perimter Institute for their hospitality while this work was done. The work of M.B.W. was supported in part by the U.S. Department of Energy under contract No. DE-FG02-92ER40701. 

\section{Appendix: EWPD calculations}

The one loop results for the $S_8$ model are the same as for the model
discussed in \cite{Manohar:2006ga,Burgess:2009wm}. We use the $\rm STUVWX$ parameterization \cite{Maksymyk:1993zm} of 
EWPD as the mass scale of the new scalars is $\sim 100 \, \rm GeV$. The relevant results in terms of 
Passarino-Veltman functions \cite{Passarino:1978jh} with standard definitions\footnote{With $c,s$ the cosine and sine of the weak mixing angle.} are

\begin{eqnarray}
\delta \Pi_{WW}(p^2) &=& \frac{g_1^2}{2 \pi^2}\Big[   B_{22}(p^2,m_I^2,m_+^2) \nn
+  B_{22}(p^2,m_R^2,m_+^2),\\ &-& \frac{1}{2} A_0(m_+^2)-\frac{1}{4} A_0(m_R^2)-\frac{1}{4} A_0(m_I^2)\Big], \nn \\
\delta \Pi_{ZZ}(p^2) &=& \frac{g_1^2 }{2 \pi^2 c^2}\Big[  
(1-2s^2)^2 \left( B_{22}(p^2,m_+^2,m_+^2) -\frac{1}{2} A_0(m_+^2)\right), \nn \\
&+&  B_{22}(p^2,m_R^2,m_I^2 )-\frac{1}{4} A_0(m_R^2)-\frac{1}{4} A_0(m_I^2)\Big], \nn \\
\delta \Pi_{\gamma\gamma}(p^2) &=& \frac{2 e^2 }{ \pi^2 }\Big[  
 B_{22}(p^2,m_+^2,m_+^2) -\frac{1}{2} A_0(m_+^2) \Big], \nn \\
\delta \Pi_{\gamma Z}(p^2) &=& \frac{e g_1(1-2s^2)}{ \pi^2 c }\Big[  
 B_{22}(p^2,m_+^2,m_+^2) -\frac{1}{2} A_0(m_+^2) \Big]. \nn
\end{eqnarray}

For $p^2=0$ these expressions become
\begin{eqnarray}
\delta \Pi_{WW}(0) &=& \frac{g_1^2 }{8 \pi^2} \left( \frac{1}{2} f(m_+,m_R) +  \frac{1}{2} f(m_+,m_I) \right), \nn \\
\delta \Pi_{ZZ}(0) &=& \frac{g_1^2}{8 \pi^2 c^2} \left( \frac{1}{2} f(m_R,m_I) \right),  \nn
\end{eqnarray}
where 
\begin{equation}
 f(m_1,m_2)=m_1^2 +m_2^2 -\frac{2 m_1^2 m_2^2}{m_1^2-m_2^2}\log{ \frac{m_1^2}{m_2^2} }. \nn
\end{equation}

The derivatives of the vacuum polarizations are
\begin{eqnarray}
\delta \Pi'_{\gamma\gamma}(0)&=& - \frac{e^2}{6 \pi^2} B_0(0,m_+^2,m_+^2), \nn \\
\delta \Pi'_{\gamma Z}(0) &=& -\frac{e g_1(1-2s^2)}{12 \pi^2 c } B_0(0,m_+^2,m_+^2), \nn\\
\delta \Pi'_{WW}(p^2) &=& \frac{g_1^2}{2 \pi^2} \Big[ -\frac{1}{6}\Delta + \frac{\partial b_{22}(p^2,m_I^2,m_+^2)}{\partial p^2} , \nn \\ 
&\,& \hspace{1.9cm} + \frac{\partial b_{22}(p^2,m_R^2,m_+^2)}{\partial p^2} \Big], \nn \\
\delta \Pi'_{ZZ}(p^2) &=& \frac{g_1^2}{2 \pi^2 c^2}\Big[  
-  \frac{1}{12}\Delta +\frac{ \partial b_{22}(p^2,m_R^2,m_I^2 )}{\partial p^2}, \nn \\
&\,& (1-2s^2)^2 \left( -\frac{1}{12}\Delta +\frac{\partial b_{22}(p^2,m_+^2,m_+^2)}{\partial p^2} \right) \Big]. \nn 
\end{eqnarray}

Using these results we can construct the $\rm STUVWX$ parameters
with the standard definitions  \cite{Maksymyk:1993zm}

\begin{eqnarray}
 \frac{\alpha S}{4 s^2 \, c^2} &=& \left[\frac{\delta \Pi_{ZZ}(M_Z^2) - \delta \Pi_{ZZ}(0)}{M_Z^2} \right], \nn \\
  &\,&- \frac{(c^2 - s^2)}{s \, c}  \delta \Pi'_{Z\, \gamma}(0)  -  \delta \Pi'_{\gamma \, \gamma}(0), \nn \\
\alpha T &=& \frac{\delta \Pi_{WW}(0)}{M_W^2} - \frac{\delta \Pi_{ZZ}(0)}{M_Z^2}, \nn \\
 \frac{\alpha U}{4 s^2} &=& \left[\frac{\delta \Pi_{WW}(M_W^2) - \delta \Pi_{WW}(0)}{M_W^2} \right], \nn \\
&\,& - c^2 \left[\frac{\delta \Pi_{ZZ}(M_Z^2) - \delta \Pi_{ZZ}(0)}{M_Z^2} \right], \nn \\
&\,& - s^2 \, \delta \Pi'_{\gamma\, \gamma}(0)  -  2 \, s \, c \, \delta \Pi'_{Z \, \gamma}(0), \nn \\
  \alpha V &=&  \delta \Pi'_{Z Z}(M_Z^2) - \left[\frac{\delta \Pi_{ZZ}(M_Z^2) - \delta \Pi_{ZZ}(0)}{M_Z^2} \right], \nn \\
  \alpha W &=&  \delta \Pi'_{WW}(M_W^2) - \left[\frac{\delta \Pi_{WW}(M_W^2) - \delta \Pi_{ZZ}(0)}{M_W^2} \right], \nn \\
  \alpha X &=&  - s \, c \left[ \frac{\delta \Pi_{Z \, \gamma}(M_Z^2)}{M_Z^2} -  \delta \Pi'_{Z \, \gamma}(0) \right]. \nn
\end{eqnarray}

Here $\Delta$ is the divergence that cancels in the pseudo-observables $\rm STUVWX$ but we note we calculate in dimensional regularization and $\rm \overline{MS}$ in $d= 4 - 2 \epsilon$ dimensions.
As the number of degrees of freedom in this $S_8$ model and in the model  \cite{Manohar:2006ga}
are the same, we can directly use the detailed fit results on the allowed masses (determined from these formulas) presented in \cite{Burgess:2009wm}.
These results generally allow masses for fixed $m_h = 96^{+29}_{-24} \, {\rm GeV}$ when $m_S \gtrsim 100 \, {\rm GeV}$ characterized by
$|m_I - m_\pm| < 50 \, {\rm GeV}$ for $S_8$.


\newpage

\begin{thebibliography}{99}

\bibitem{Abazov:2010hv}
  V.~M.~Abazov {\it et al.}  [D0 Collaboration],
  arXiv:1005.2757.

\bibitem{Lenz:2010gu}
  A.~Lenz {\it et al.},
  arXiv:1008.1593.
  
\bibitem{Ikado:2006un}
  K.~Ikado {\it et al.}  [Belle Collaboration],
  Phys.\ Rev.\ Lett.\  {\bf 97}, 251802 (2006)
  [arXiv:hep-ex/0604018].

\bibitem{:2008ch}
  I.~Adachi {\it et al.}  [Belle Collaboration],
  arXiv:0809.3834.
  
\bibitem{Aubert:2007xj}
  B.~Aubert {\it et al.}  [BABAR Collaboration],
  Phys.\ Rev.\  D {\bf 77}, 011107 (2008)
  [arXiv:0708.2260 [hep-ex]].
  
\bibitem{:2008gx}
  B.~Aubert {\it et al.}  [BABAR Collaboration],
  Phys.\ Rev.\  D {\bf 81}, 051101 (2010)
  [arXiv:0809.4027 [hep-ex]].
  
 \bibitem{Aaltonen:2007he}
  T.~Aaltonen {\it et al.}  [CDF Collaboration],
  Phys.\ Rev.\ Lett.\  {\bf 100}, 161802 (2008)
  [arXiv:0712.2397 [hep-ex]].

\bibitem{:2008fj}
  V.~M.~Abazov {\it et al.}  [D0 Collaboration],
  Phys.\ Rev.\ Lett.\  {\bf 101}, 241801 (2008)
  [arXiv:0802.2255 [hep-ex]].
  
\bibitem{Lunghi:2008aa}
  E.~Lunghi and A.~Soni,
  Phys.\ Lett.\  B {\bf 666}, 162 (2008)
  [arXiv:0803.4340 [hep-ph]].

  
\bibitem{Ligeti:2010ia}
  Z.~Ligeti, M.~Papucci, G.~Perez and J.~Zupan,
  arXiv:1006.0432 [hep-ph].

 
\bibitem{Blum:2010mj}
  K.~Blum, Y.~Hochberg and Y.~Nir,
  arXiv:1007.1872.

 
\bibitem{Batell:2010qw}
  B.~Batell and M.~Pospelov,
  arXiv:1006.2127 [hep-ph].

  \bibitem{Chivukula:1987py}
  R.~S.~Chivukula and H.~Georgi,
  ``Composite Technicolor Standard Model,''
  Phys.\ Lett.\  B {\bf 188} (1987) 99.


\bibitem{Hall:1990ac}
  L.~J.~Hall and L.~Randall,
  ``Weak scale effective supersymmetry,''
  Phys.\ Rev.\ Lett.\  {\bf 65}, 2939 (1990).



  \bibitem{D'Ambrosio:2002ex}
  G.~D'Ambrosio, G.~F.~Giudice, G.~Isidori and A.~Strumia,
  Nucl.\ Phys.\  B {\bf 645} (2002) 155
  [arXiv:hep-ph/0207036].

\bibitem{Dighe:2010nj}
  A.~Dighe, A.~Kundu and S.~Nandi,
  Phys.\ Rev.\  D {\bf 82}, 031502 (2010)
  [arXiv:1005.4051 [hep-ph]].


\bibitem{Bauer:2010dg}
  C.~W.~Bauer and N.~D.~Dunn,
  arXiv:1006.1629 [hep-ph].


\bibitem{Grossman:2009mn}
  Y.~Grossman, Y.~Nir and G.~Perez,
  Phys.\ Rev.\ Lett.\  {\bf 103}, 071602 (2009)
  [arXiv:0904.0305 [hep-ph]].

  \bibitem{Demir:2002gg}
  D.~A.~Demir, M.~Pospelov and A.~Ritz,
  Phys.\ Rev.\  D {\bf 67}, 015007 (2003)
  [arXiv:hep-ph/0208257].

  
  \bibitem{Bagger:1997gg}
  J.~A.~Bagger, K.~T.~Matchev and R.~J.~Zhang,
  Phys.\ Lett.\  B {\bf 412}, 77 (1997)
  [arXiv:hep-ph/9707225].
  
  \bibitem{Dobrescu:2010rh}
  B.~A.~Dobrescu, P.~J.~Fox and A.~Martin,
  Phys.\ Rev.\ Lett.\  {\bf 105}, 041801 (2010)
  [arXiv:1005.4238 [hep-ph]].
  
\bibitem{Buras:2010mh}
  A.~J.~Buras, M.~V.~Carlucci, S.~Gori and G.~Isidori,
  arXiv:1005.5310 [hep-ph].

 \bibitem{Buras:2010zm}
  A.~J.~Buras, G.~Isidori and P.~Paradisi,
  arXiv:1007.5291 [hep-ph].
  
\bibitem{Pich:2009sp}
  A.~Pich and P.~Tuzon,
  Phys.\ Rev.\  D {\bf 80}, 091702 (2009)
  [arXiv:0908.1554 [hep-ph]].
    
\bibitem{Jung:2010ik}
  M.~Jung, A.~Pich and P.~Tuzon,
  arXiv:1006.0470 [hep-ph].

  
  
  \bibitem{Feldmann:2008ja}
  T.~Feldmann and T.~Mannel,
  Phys.\ Rev.\ Lett.\  {\bf 100} (2008) 171601
  [arXiv:0801.1802 [hep-ph]].

  \bibitem{Kagan:2009bn}
  A.~L.~Kagan, G.~Perez, T.~Volansky and J.~Zupan,
  arXiv:0903.1794 [hep-ph].

\bibitem{Manohar:1983md}
  A.~Manohar and H.~Georgi,
  Nucl.\ Phys.\  B {\bf 234}, 189 (1984).

 
\bibitem{Boyd:1990bx}
  G.~Boyd, A.~K.~Gupta, S.~P.~Trivedi and M.~B.~Wise,
  Phys.\ Lett.\  B {\bf 241}, 584 (1990).

  \bibitem{Braaten:1990gq}
  E.~Braaten, C.~S.~Li and T.~C.~Yuan,
  Phys.\ Rev.\ Lett.\  {\bf 64}, 1709 (1990).
   
   \bibitem{Weinberg:1989dx}
  S.~Weinberg,
  Phys.\ Rev.\ Lett.\  {\bf 63}, 2333 (1989).

  
   
\bibitem{Manohar:2006ga}
  A.~V.~Manohar and M.~B.~Wise,
  Phys.\ Rev.\  D {\bf 74}, 035009 (2006)
  [arXiv:hep-ph/0606172].

  
  \bibitem{Grzadkowski:2008mf}
  B.~Grzadkowski and M.~Misiak,
  Phys.\ Rev.\  D {\bf 78}, 077501 (2008)
  [arXiv:0802.1413 [hep-ph]].

  \bibitem{Barberio:2007cr}
  E.~Barberio {\it et al.}  [Heavy Flavor Averaging Group (HFAG)
                  Collaboration],
  arXiv:0704.3575 [hep-ex].



\bibitem{Burgess:2009wm}
  C.~P.~Burgess, M.~Trott and S.~Zuberi,
  JHEP {\bf 0909}, 082 (2009)
  [arXiv:0907.2696 [hep-ph]].
  
  
  \bibitem{Martin:2009iq}
  A.~D.~Martin, W.~J.~Stirling, R.~S.~Thorne and G.~Watt,
  Eur.\ Phys.\ J.\  C {\bf 63}, 189 (2009)
  [arXiv:0901.0002 [hep-ph]].

  
  
\bibitem{Dawson:2005vi}
S.~Dawson, C.~B. Jackson, L.~Reina, and D.~Wackeroth,
\newblock Mod. Phys. Lett. {\bf A21}, 89 (2006), hep-ph/0508293.
  

\bibitem{Mantry:2007ar}
  S.~Mantry, M.~Trott and M.~B.~Wise,
  Phys.\ Rev.\  D {\bf 77}, 013006 (2008)
  [arXiv:0709.1505 [hep-ph]].


\bibitem{Glashow:1976nt}
  S.~L.~Glashow and S.~Weinberg,
  Phys.\ Rev.\  D {\bf 15}, 1958 (1977).
      
\bibitem{Arnold:2009ay}
  J.~M.~Arnold, M.~Pospelov, M.~Trott and M.~B.~Wise,
  JHEP {\bf 1001}, 073 (2010)
  [arXiv:0911.2225 [hep-ph]].



  \bibitem{Peskin:2001rw}
  M.~E.~Peskin and J.~D.~Wells,
  Phys.\ Rev.\  D {\bf 64}, 093003 (2001)
  [arXiv:hep-ph/0101342].
            
\bibitem{Passarino:1978jh}
  G.~Passarino and M.~J.~G.~Veltman,
  Nucl.\ Phys.\  B {\bf 160}, 151 (1979).

\bibitem{Maksymyk:1993zm}
  I.~Maksymyk, C.~P.~Burgess and D.~London,
  Phys.\ Rev.\  D {\bf 50}, 529 (1994)
  [arXiv:hep-ph/9306267].


 \end{thebibliography}
\end{document}